\documentclass[
superscriptaddress, twocolumn,
 amsmath,amssymb,
 aps,
 pre,
floatfix,
]{revtex4-1}

\usepackage{graphicx}
\usepackage{bm}
\usepackage{amsmath}
\usepackage{mathrsfs}

\begin{document}


\title{Geometry-induced circulation of local photonic transport in a triangular metastructure}

\author{Palak Dugar}
\affiliation{Department of physics, University of California, Merced, CA 95343, USA.}
\author{Michael Scheibner}
\affiliation{Department of physics, University of California, Merced, CA 95343, USA.}
\author{Chih-Chun Chien}%
\email{cchien5@ucmerced.edu}
\affiliation{Department of physics, University of California, Merced, CA 95343, USA.}


\date{\today}

\begin{abstract}
A geometry-based mechanism for inducing circulation of photons is illustrated by a metastructure consisting of quantum dots arranged in a triangle coupled to photonic structures. The coupling between the photons and the excitons in the quantum dots leads to a photon blockade and limits the number of photons participating in the transport. In the steady state described by the quantum master equation of photons, the local photonic currents exhibit distinct circulation patterns, which originate from the wave nature in a multi-path geometry. The geometry-based mechanism does not require an artificial gauge field from light-matter interactions. The phase diagrams showing where different patterns of circulation can be found saturate as the number of photons allowed on each site increases. By using the third-quantization formalism, we show the circulation survives without any photon blockade in the noninteracting case. Moreover, we demonstrate the decoupling of the direction of the local current from the density difference and propose possible applications of the local photonic transport.
\end{abstract}

\maketitle

\section{\label{sec:level1}Introduction} 
Quantum systems with interesting photonic transport properties play an instrumental role in quantum simulations \cite{hartmann2016quantum, houck2012chip} and quantum information \cite{peyronel2012quantum,brandes2005coherent,northup2014quantum}. It is possible to induce effective photon interactions using optical nonlinearities in the medium \cite{hartmann2016quantum}, so photonic systems can be used to simulate a variety of model Hamiltonians such as the Bose-Hubbard model \cite{longhi2011optical}, Jaynes-Cummings-Hubbard model \cite{rossini2007mott}, and Lieb-Lininger model \cite{chang2008crystallization}. 
On the other hand, photons are noninteracting in the absence of medium effects, so they provide an inherently decoherence-free system. By coupling the system with photonic reservoirs, the cavity- or circuit- quantum electrodynamics provides an ideal ground for simulating open quantum systems \cite{houck2012chip, rai2015photonic, schmidt2010nonequilibrium}. Meanwhile, the rapid propagation of photons \cite{flamini2018photonic} and the ability to be highly stable carriers of quantum information \cite{northup2014quantum} make photons an important player in quantum information technology.

Photonic transport is a non-equilibrium process, which may be described by the open quantum system approach~\cite{breuer2002theory} or other frameworks~\cite{aeberhard2011theory, sieberer2016keldysh, kordas2015dissipative, prosen2014boundary,thingna2012steady}.
There has been recent interest in coupling photonic-crystal arrays and quantum dots for enhanced efficiency of the emission rate of the photons from the quantum dots \cite{chu2019lifetimes}, as well as exciting physics such as coherent superradiance and high fidelity in quantum emitters for quantum network applications~\cite{grim2019scalable, atature2018material}. 
Meanwhile, geometry plays an important role in quantum transport. For instance, the propagation speed of a wavepacket reflects the group velocity, which depends on the underlying lattice structure~\cite{metcalf2016matter}. Moreover, counter-intuitive local flows can arise due to multi-path geometries. For example, Ref. \cite{arnold2013magnetic} shows the dynamics of electrons in a conducting ring embedded in a cavity coupled to electrically biased leads may exhibit a persistent circular current in the presence of a constant magnetic field.
Ref. \cite{renou2019genuine} shows that non-local correlations can be achieved by entangling the photons placed in a triangle. In absence of any magnetic field, Ref. \cite{lai2018tunable} shows that the steady-state electronic current flowing through a triangular triple quantum dot may exhibit local circulations and reasons that the circulation is due to the wave nature of the electrons spreading across the underlying multi-path geometry. Ref.~\cite{dugar2019geometry} shows that local thermal current may flow from cold to hot in classical harmonic systems by applying the idea of transporting waves in multi-path geometries.

There are many ways to induce a circulating current. For charged particles like the electrons, an external magnetic field can induce a circulating current in classical~\cite{griffiths1999introduction} as well as quantum~\cite{Jayich09} systems.
Although photons are charge neutral, it is possible to exploit light-matter interactions to generate an artificial gauge field \cite{hey2018advances, fang2012photonic,reitzenstein2010exciton}, thereby driving the photons the way a magnetic field drives a charged particle. Instead, here we explore the geometry-induced circulation, which has been shown to work for electrons and classical harmonic systems \cite{lai2018tunable, dugar2019geometry}. As we will show shortly, the geometry-based mechanism works for photonic transport and establishes its universal properties. 

\begin{figure}[th]
	\centering
	\includegraphics[width=0.49\textwidth]{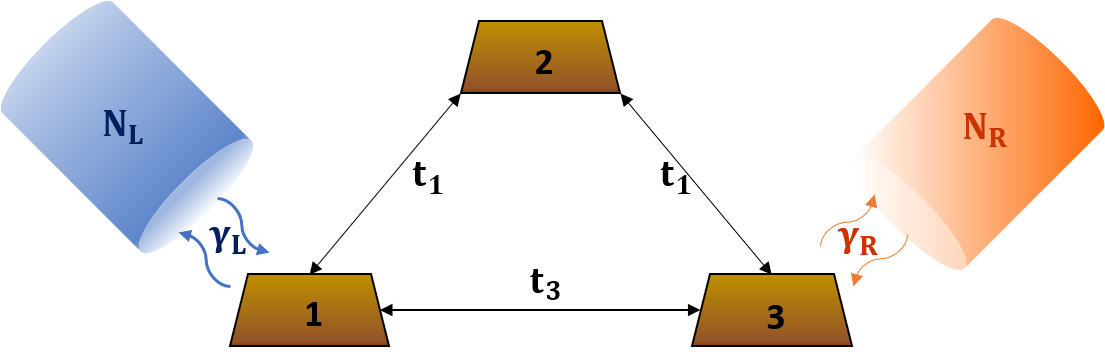}
	\caption{\label{fig:schematic} The quantum dot - photonics metastructure for studying geometry-induced photon circulation. The three quantum dots, labeled by $1$, $2$, and $3$, are embedded in photonic structures, which are connected by photonic waveguides. Site $1$ ($3$) is connected to a photon pump (sink). Only the photons coupled to the excitons in the quantum dots are transported between the sites.}
\end{figure}

There have been many theoretical frameworks for studying photonic systems coupled to reservoirs~\cite{mertz2016photonic, sieberer2016keldysh, kordas2015non, pivzorn2013one}. Ref. \cite{sieberer2016keldysh} reviews the open system Keldysh functional integral approach and its applications while 
Refs. \cite{mertz2016photonic, kordas2015non, pivzorn2013one} used different methods to study photons modeled by the Bose-Hubbard model.
In the following, the dynamics of the photons will be described by the Lindblad equation \cite{breuer2002theory}, which allows a detailed analysis of the local current through each link in a multi-path geometry. We will consider a metastructure realizable by integrating three quantum dots with photonic structures and waveguides, as illustrated in Fig.~\ref{fig:schematic}. The system is coupled to two photonic reservoirs for sustaining a steady-state current. The photonic transport in the metastructure will be shown to exhibit steady-state circulations without any artificial gauge field. Importantly, the orientation and magnitude of the circulation can be tuned by the system parameters as well as the system-reservoir coupling.

While the Bose statistics of photons allows arbitrary numbers of photons on a given site, the quantum dot - photonics metastructure considered here has a constraint on the photons that can participate in transport. Due to the coupling between the photons and the excitons in the quantum dots, the spectrum of the coupled photons may differ from that of the uncoupled photons. With carefully designed waveguides, only the photons coupled to the excitons can be transmitted~\cite{DuttaReview}. Given the limited number of excitons on each quantum dot, the number of photons on each site that can participate in transport is limited.  This phenomenon is generally known as the photon blockade \cite{kyriienko2019nonlinear,kasprzak2006bose,verger2006polariton}.
Tuning the gate voltage of the quantum dot may change the number of excitons, thereby providing a possible way to control the number of photons on each site. 

The ability to control the local photonic current may find future applications. By encoding the binaries $0$ and $1$ by the orientations of the circulation, the photonics - quantum dot mestastructure may serve as a memory element similar to the one uses only the electronic transport in triangular triple quantum dots~\cite{lai2018tunable}. Moreover, we will show that the steady-state photonic transport establishes the decoupling of the flow direction and the density difference. Thus, it is possible to design systems which can transport photons in a preferential direction regardless of the number of photons on the sites.

The rest of the paper is organized as follows.  Section~\ref{sec:level2} describes the metastructure and its modeling. The effective Hamiltonian and the Lindblad quantum master equation for studying the photonic transport are also presented. Section~\ref{sec:level3a} shows the numerical results for the noninteracting case, where we identify the clockwise and counterclockwise circulations. In section~\ref{sec:level3b}, we present the phase diagrams of the metastructure in the weak and strong interaction regime and discuss how the interaction and system parameters influence the photon circulations. Possible applications and experimental realization of the system are discussed in Sections~\ref{sec:level3c} and ~\ref{sec:level3d}. Section~\ref{sec:level3e} proves that the photon blockade is not essential for the circulation by showing the results without any limitation on the photon number.
We conclude our work in Section~\ref{sec:level4}. The Appendix summarizes the third-quantization formalism for the system without photon blockade.

\section{\label{sec:level2}Model and Method}
The system considered here consists of three quantum dots in a triangular geometry coupled to photonic structures, which allow photons to be transported across the sites. As illustrated in Fig. \ref{fig:schematic}, the quantum dot - photonics metastructure is connected to two photon reservoirs acting as incoherent photon pump and sink for sustaining the photonic transport. 
We treat the metastructure as an open quantum system driven by the two reservoirs. Physical observables, such as the density or current, can be obtained from the corresponding expectation values once the time evolution of the reduced density matrix of the system, $\rho(T)$, is known. In general, the evolution of $\rho$ is not unitary under the influence of the reservoirs and cannot be described by the Liouville - von Neumann equation. Such a non-unitary evolution of $\rho$, nevertheless, can be investigated by the Lindblad quantum master equation of the form~\cite{breuer2002theory,haroche2006exploring,lai2018tunable}:
\begin{eqnarray}\label{eq:LQME}
\frac{\partial\rho(T)}{\partial T}&=&   \mathscr{L}\rho(T) \nonumber \\
&=&\frac{\iota}{\hbar}[\rho,\mathscr{H}]+ \gamma_{L}N_{L}(c_{1}^{\dagger}\rho c_{1} -\frac{1}{2}\{c_{1}c_{1}^{\dagger},\rho\})  \nonumber \\ 
& &+ \gamma_{L}(N_{L}+1)(c_{1}\rho c_{1}^{\dagger} -\frac{1}{2}\{c_{1}^{\dagger}c_{1},\rho\})  \nonumber \\ 
& &+ \gamma_{R}N_{R}(c_{3}^{\dagger}\rho c_{3} -\frac{1}{2}\{c_{3}c_{3}^{\dagger},\rho\})  \nonumber \\ 
& &+ \gamma_{R}(N_{R}+1)(c_{3}\rho c_{3}^{\dagger} -\frac{1}{2}\{c_{3}^{\dagger}c_{3},\rho\}).
\end{eqnarray}
Here $\mathscr{L}$ is the Lindbladian, a super-operator describing the non-unitary time evolution of $\rho$.  $\mathscr{H}$ is the system Hamiltonian. $[A,B]$ and $\{A,B\}$ represent the commutator and anti-commutator of operators $A$ and $B$.  $c_{i}^{\dagger}$ and $c_{i}$ are the photonic creation and annihilation operators at site $i$. 

The left and right system-reservoir coupling constants are $\gamma_{L}$ and $\gamma_{R}$, respectively. Assuming the left and right reservoirs have fixed photon numbers $N_L$ and $N_R$, respectively, the reservoirs emit photons at the rate  $\gamma_{j}N_{j}$ into the system while it absorbs photons at the rate $\gamma_{j}(N_{j}+1)$ with $j=L, R$, as shown in Eq.~\eqref{eq:LQME}. 
Those exchange rates of photons follow the assumption of Bose statistics and instantaneous relaxation of the reservoirs. The rates also maintain the system in equilibrium if only one reservoir is connected. Moreover, the rate at which the photons are exchanged is consistent with the quantum optical master equation \cite{breuer2002theory, haroche2006exploring}. There are three basic assumptions~\cite{breuer2002theory, haroche2006exploring} behind the Lindblad equation~\eqref{eq:LQME}: (1) The Born approximation, which assumes the system interacts weakly with the reservoirs so that the influence of the interaction is negligible on the reservoirs. (2) The Markov approximation, or the memory-less reservoir condition, under which the time scale associated with the system dynamics is taken to be longer than the reservoir correlation time. (3) The secular approximation, under which one assumes the timescale associated with the system dynamics is smaller compared to the relaxation timescale of the system so that the fast oscillating terms can be discarded.

A full description of the photons interacting with the excitons in the quantum dots requires the Hamiltonian to include photon-electron and electron-electron interactions. Here we take a phenomenological point of view and model the photons in the metastructure by an effective Bose-Hubbard Hamiltonian, which includes the hopping of the photons and effective interactions between the photons due to the underlying electrons. Explicitly,
\begin{equation}\label{eq:BHH}
    \begin{split}
    \mathscr{H} & = -t_{1}(c_{1}^{\dagger}c_{2} + c_{2}^{\dagger}c_{1} + c_{2}^{\dagger}c_{3} + c_{3}^{\dagger}c_{2})\\ 
    & -t_{3}(c_{1}^{\dagger}c_{3} + c_{3}^{\dagger}c_{1}) +\frac{U}{2}\sum_{i=1}^{3}n_{i}(n_{i}-1).   
    \end{split}
\end{equation}
Here $t_{1}$ is the tunnelling coefficient between the first and second sites and between the second and third sites, $t_{3}$ is the tunneling coefficient between the first and third sites. $U$ is the the onsite coupling constant and we assume repulsive interactions with $U\ge 0$. $n_{i}=c_{i}^{\dagger}c_{i}$ is the photonic number operator at site $i$. We choose $\hbar=1$ with the time unit $T_{0}=\frac{\hbar}{t_{1}}$.

The photonic current operator from site $i$ to site $j$ is given by
\begin{equation}\label{localcurrent}
    J_{ij}=-\iota(t_{ij}c_{i}^{\dagger}c_{j}-t_{ij}c_{j}^{\dagger}c_{i}),
\end{equation}
where $t_{ij}$ takes the value $t_1$ or $t_3$ for $J_{12}$ or $J_{13}$, respectively. The  expectation value $\langle A\rangle$ of an operator $A$ can be obtained from $Tr(\rho A)$.  Here $Tr$ denotes the trace. After the steady-state density matrix $\rho_{ss}$ is found from Eq.~\eqref{eq:LQME}, the steady-state current and density can be obtained accordingly. In the steady state, $\langle J_{12}\rangle=\langle J_{23}\rangle$ if the photon loss inside the system is negligible. Thus, the total steady-state photonic current through the metastructure is given by the steady-state value of
\begin{equation}
    \langle J_{T}\rangle=\langle J_{13}\rangle+\langle J_{12}\rangle.
\end{equation}

In absence of any constraint on the photon number on each site, the density matrix for a bosonic system can be infinite dimensional because each site can accommodate any number of photons due to the Bose statistics. However, the coupling between the photons and the electrons in the quantum dots leads to a photon blockade due to the available excitons in the quantum dots. Therefore, each site can accommodate at most $M$ photons. The photon blockade thus introduces a truncated basis in the Fock space, allowing at most $M$ photons on one site. The states in the truncated Fock space can then be constructed following Ref. \cite{szabados2012efficient}.

The Lindblad equation \eqref{eq:LQME} describes a homogeneous Markov process. According to Ref. \cite{rivas2012open}, there exists at least one steady-state solution in a finite-dimensional space, which applies to the case with the photon blockade. Ref. \cite{rivas2012open} shows that when $\rho$ is rewritten as an equivalent column vector, the Lindblad superoperator $\mathscr{L}$ can be written as a square matrix. 
However, $\mathscr{L}$ is usually not of the form of a normal matrix and may not be diagonalizable. 
To overcome this technical difficulty, we implement the fourth order Runge-Kutta method \cite{press2007numerical} to integrate Eq. \eqref{eq:LQME} from a given initial state and obtain the steady-state density matrix after the transient behavior decays away. By evaluating the expectation values in the steady state, the local photonic currents and occupation numbers can be found.

In our simulations, the initial density matrix was set to the one with no photons or maximally allowed numbers of photons in the system. Both types of initial states lead to the same steady-state density matrix. However, in the strongly interacting regime when $\gamma T_{0}$ is very small, the convergence to the steady-state value can be slow, so we use the steady-state value of an adjacent point in the parameter space as the initial condition to achieve faster convergence. Importantly, we have checked that $\rho_{ss}$ obtained from the numerical integration of Eq.~\eqref{eq:LQME} indeed makes the right-hand side vanish within machine precision.

\begin{figure}[t]
	\centering
	\includegraphics[width=0.49\textwidth]{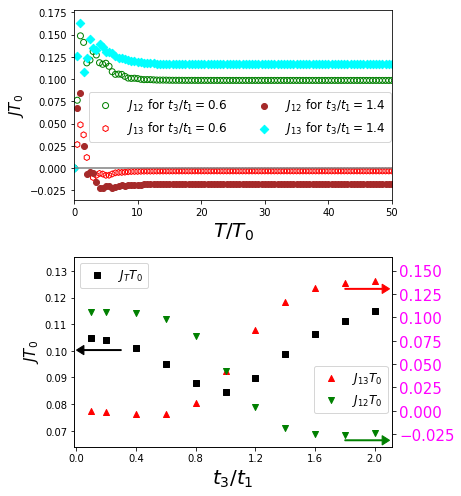}
	\caption{\label{fig:M_1} (Top panel) The local currents $J_{12}$ and $J_{13}$ as functions of time with $t_3/t_1=0.6,1.4$, $\gamma T_{0}=0.5$, and $M=1$ without the onsite interaction ($U=0$). The plateaus of the currents are the signature of a steady state. (Bottom panel) The steady-state values of the local currents $J_{12}, J_{13}$ and the total current $J_{T}$ as functions of $t_3/t_1$ with the same $\gamma T_{0}$ and $M$, showing opposite signs of $J_{12}$ and $J_{13}$ in certain regimes.}
\end{figure}

\begin{figure*}[t]
	\centering
	\includegraphics[width=0.95\textwidth]{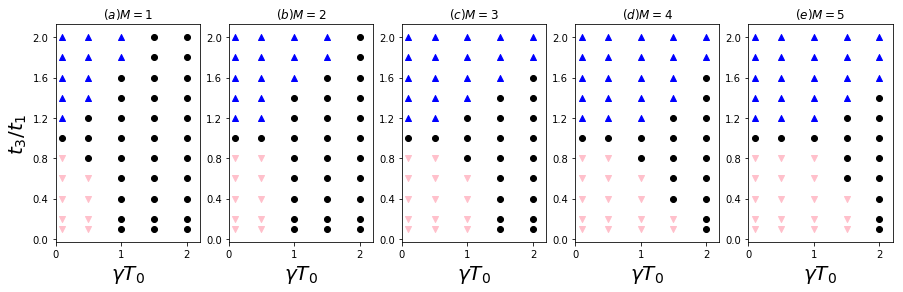}
	\caption{\label{fig:U_0} Phase diagrams showing the  steady-state patterns of the photonic current with $U=0$ and $M=1,2,3,4,5$ (from left to right). Here the pink upside-down triangles denote the CW circulation, the black circles denote the UD flow, and the blue triangle denote the CCW circulation.}
\end{figure*}

\section{\label{sec:level3}Results and discussions}

\subsection{\label{sec:level3a}Noninteracting photons with photon blockade}
We begin with the photonic transport through the metastructure illustrated in Fig.~\ref{fig:schematic} in the incoherently driven-dissipative regime with $\gamma_{L}=\gamma_{R}=\gamma\neq 0$ and fixed $N_{L}=1$ and $N_{R}=0$. The photon blockade due to the photon-exciton interaction will cap the number of photons on each site by $M$. We begin with the noninteracting Hamiltonian with $U=0$ and discuss the interaction effect later.

The case with at most one photon per site, $M=1$, may be experimentally realized by tuning the gate voltages to allow only one exciton per quantum dot.
The upper panel of Fig. \ref{fig:M_1}
shows the time dependence of the local currents for $t_3/t_1=0.6,1.4$ and $\gamma T_{0}=0.5$, exhibiting steady-state behavior in the long-time limit. Importantly, one can see that $J_{13}$ of the case with $t_3/t_1=0.6$ as well as $J_{12}$ of the case with $t_3/t_1=1.4$ flow against the pumping of the reservoirs.

The steady-state values of the local and total currents are then extracted for different values of $t_3/t_1$ with fixed $\gamma T_{0}=0.5$ and shown in the lower panel of Fig. \ref{fig:M_1}.
Note that the combination of $J_{12}>0$ and $J_{13}<0$ corresponds to a clockwise (CW) circulation of the photons in the metastructure shown in Fig.~\ref{fig:schematic} while the combination of $J_{12}<0$ and $J_{13}>0$ corresponds to a counter-clockwise (CCW) circulation. There are other points showing both $J_{12}>0$ and $J_{13}>0$, corresponding to the unidirectional (UD) flow. Throughout the paper, we will use this criterion for labeling the orientation of the circulation. As the ratio $t_3/t_1$ changes, the internal circulation of the photons in the steady state can be tuned.

In the bottom panel of Fig. \ref{fig:M_1}, the photons show CW circulation (UD flow) when $t_{3}/t_{1}$ is small (intermediate in the range $0.8 \leq t_{3}/t_{1} \leq 1.2$). For large $t_{3}/t_{1}$, the photons exhibit CCW circulation. We emphasize that the internal circulation of photons is a steady-state phenomenon, not a transient one, because the long-time limit has been taken.
The mechanism behind the photonic circulations is a combination of the wave nature of quantum particles and multi-path geometry. As explained in Ref. \cite{lai2018tunable}, the wavefunctions spread out over the whole system in the triangular geometry during the dynamic process, making it possible for one path to overflow while another path transports the particles backward to compensate for it. Utilizing the wave nature and multi-path geometry, Ref.~\cite{dugar2019geometry} shows that local thermal current can flow in the direction opposite to the total current as well.
Since the mechanism transcends spin-statistics, here we found circulation of photons in a similar setup.  Nevertheless, the Bose statistics of photons allows us to explore the dependence of the circulation on the number of particles allowed on each site, and we will address this issue shortly. Moreover, the geometry-induced circulation is not associated with any real or artificial gauge field. Thus, there is no quantization condition on the vorticity of the photons.

In the $M=1$ photon-blockade case, the non-monotonic dependence of $J_{T}$ on $t_3/t_1$ shown in Fig. \ref{fig:M_1} may be considered as an indication of a change of the orientation of the photon circulation in the metastructure.  
We notice that the $M=1$ case of photons is similar to the spinless fermions studied in Ref.~\cite{lai2018tunable} because there cannot be two particles on the same site. However, a careful comparison shows that the similarity is only qualitative because of the different spin-statistics. Specifically, the evaluations of the commutator in Eq.~\eqref{eq:LQME}  and the exchange terms with the reservoirs depend on whether bosons or fermions are considered. The spin-statistics thus causes quantitative differences between the  electronic transport and the photonic transport with the constraint $M=1$.

After establishing the existence of internal photon circulations in the metastructure, it is important to check if the circulation can survive in the photon blockade regime with more photons per site, i.e., when $M>1$. In the following, we consider the cases with $M=2,3,4,5$ and show that the phase diagrams of the photon circulation do not change much as $M$ increases. Later on we will show the phase diagram without any photon blockade,  $M\rightarrow\infty$, and confirm that the photonic transport already saturates for relatively small values of $M$.

Fig. \ref{fig:U_0} shows the phase diagrams of the steady-state flow patterns as a function of $M$, $t_3/t_1$, and $\gamma$. By comparing the phase diagrams of noninteracting photons with $M=1,2,3,4,5$, we infer that in the photon blockade regime with higher-numbers of photons per site, the internal circulations of photons survive and the regimes of both CW and CCW circulations increase with $M$. As will be shown later, more photons are present in each site when $M$ increases. This in turn increases the possible configurations of the photon wavefunctions, leading to overshoots of the flows along certain paths and causing the internal circulation.  

\begin{figure*}[t]
	\centering
	\includegraphics[width=0.99\textwidth]{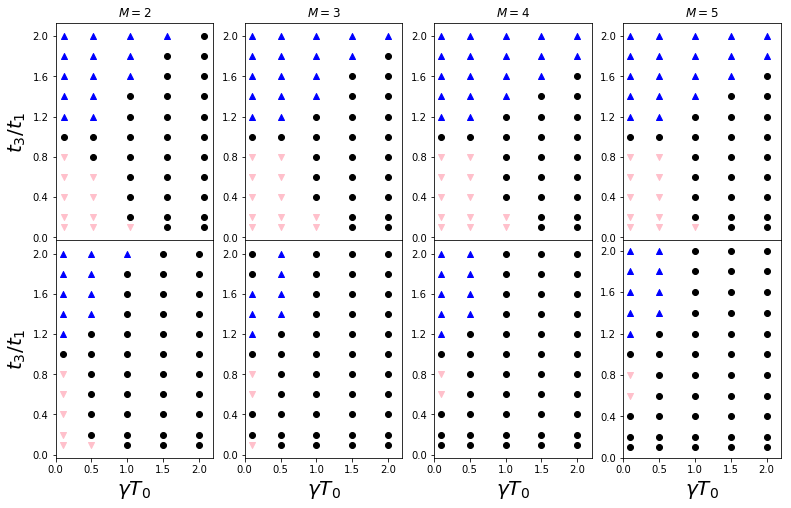}
	\caption{\label{fig:U_no_0} 
	Phase diagrams showing the steady-state patterns of the photonic current for the cases with $U=t_1$ (the top row) and $U=5t_1$ (the bottom row). Here $M=2, 3, 4, 5$ from left to right.	The pink upside-down triangles denote the CW circulation, the black circles denote the UD flow, and the blue triangles denote the CCW circulation.}
\end{figure*}

\subsection{\label{sec:level3b}Photons with effective interactions and photon blockade}
While photons in vacuum do not have bare interactions with each other~\cite{roychoudhuri2008nature}, the photons in the metastructure shown in Fig.~\ref{fig:schematic} interact with the excitons in the quantum dots and may experience an effective repulsion among themselves. To investigate photonic transport in the presence of the effective interactions, we introduce an effective onsite interaction to Eq.~(\ref{eq:BHH}) and calculate the steady-state expectation values of the local currents from Eq.~\eqref{eq:LQME}. After extracting the steady-state pattern of the local flow, the phase diagrams of the interacting photons in the metastructure are shown in
Fig. \ref{fig:U_no_0}. 

When the coupling constant $U$ is smaller or comparable to the hopping coefficients, the results shown in the upper row of Fig.~\ref{fig:U_no_0} are qualitatively similar to the non-interacting cases shown in Fig.~\ref{fig:U_0}. The regimes showing CW and CCW circulations increase as $M$ increase from $1$ to $4$, but then the circulation regimes seem to saturate as $M$ further increases. The results thus establish that the photon circulation is not unique to noninteracting systems.

As $U$ increases further, the regimes showing CW or CCW circulations are suppressed, as shown
in the lower panel of Fig.~\ref{fig:U_no_0}. For electronic transport in a triangular triple quantum dot system, a similar suppression of circulations by onsite interactions has been discussed in Ref. \cite{lai2018tunable}, where scattering of the electrons due to the onsite interactions is believed to be the reason for the suppression. Here the effective scattering of photons may suppress the circulation as well, but the Bose statistics of photons leads to more complicated phase diagrams as shown in the bottom row of Fig.~\ref{fig:U_no_0}.

The regime of weak $\gamma$ and strong $U/t_1$ is interesting in the sense that as $t_{3}/t_{1}$ increases, intermediate regimes with CW or CCW circulation are enclosed by the unidirectional-flow regime for $M=3,4,5$. For example, The lower-left parts of the phase diagrams shown in the bottom row of Fig.~\ref{fig:U_no_0} have a CW regime emerging at small values of $t_3/t_1$, but the UD regime occurs both above and below the CW regime. The competition among the different circulations in the small $\gamma$ and strong interaction regime implies the system is sensitive to the parameters, adding challenges to an accurate measurement of the phase diagrams of strongly interacting non-equilibrium systems.

\begin{figure}[t]
	\centering
	\includegraphics[width=0.45\textwidth,  height=0.75\textwidth]{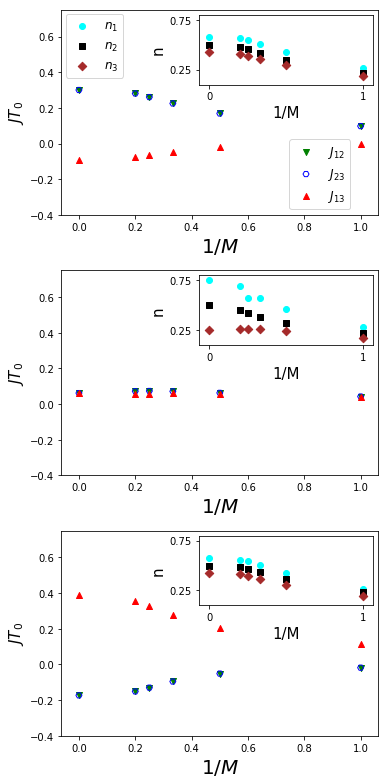}
	\caption{\label{fig:1_by_M} The dependence of the local steady-state currents $J_{12}$ (blue hollow hexagons), $J_{23}$ (green triangles), and $J_{13}$ (red upside-down triangles) on $1/M$ for $t_{3}/t_{1}=0.6,1.0,1.4$ from top to bottom. Here $\gamma T_{0}=0.5$ and $U=0$. $J_{12}$ and $J_{23}$ overlap in the steady state. The insets show the photon occupation numbers on the three sites, $n_1, n_2$, and $n_3$ (cyan circles, black squares, and brown diamonds), as functions of $1/M$ with the same parameters as those in the main panels. }
\end{figure}

\subsection{\label{sec:level3c}Experimental implications}
Refs. \cite{stegmann2020current, bellec2013tight} present experimental probes of transport behavior based on the similarity between the Helmholtz equation of electromagnetic waves and the Schrodinger equation of quantum particles. The role of the local current of single-particle quantum mechanics is played by the transmission coefficient of microwaves between adjacent macroscopic resonators. In particular, Ref. \cite{stegmann2020current} experimentally verifies that the transmission coefficients of the microwave correspond to a circulating current in a benzene-like hexagon, showing the wave nature and multi-path geometry are able to induce an internal circulation. While the microwave simulators have demonstrated circulating current for single-particle transport of the analogous quantum systems, the metastructure presented here will offer a route to the study of many-body quantum transport and establish the circulating current in the presence of interactions. In addition, cold-atom simulators may also help shed light on many-body quantum transport~\cite{ChienNatPhys}.

The photon blockade in the metastructure may be realized as follows. The photons coupled to the excitons in the metastructure have two exciton-polariton branches, an upper one and a lower one. The reservoir sends photons tuned to a frequency slightly above the lower branch to incoherently pump the photons into the metastructure, similar to the idea of Ref.~\cite{sieberer2016keldysh}. The waveguides, through which the photons can travel  \cite{lai2018tunable}, are also tuned to the frequency of the lower branch, and we focus on those coupled photons and their transport in the metastructure. 

Before presenting possible application of the geometry-induced photon circulation, we show how the local currents and densities of the photons depend on $M$ due to the photon blockade. The value of $M$ is limited by the number of excitons available for coupling to the photons on each site. As suggested in Ref.~\cite{lai2018tunable}, the number of excitons in each quantum dot may be tuned by the gate voltage. Since we consider the photonic structures that only transport the photons coupled to the excitions in the metastructure shown in Fig.~\ref{fig:schematic}, the number of photons on each site may also be tuned by the gate voltage. However, each quantum dot may not accommodate more than a few excitons, so tuning the gate voltage may be more suitable for $M\le 2$.
To study transport in a photon blockade regime with $M>2$, it is possible to adopt and modify the quantum-dot metastructure of Ref.~\cite{grim2019scalable}. The idea is to let each vertex of the triangle shown in Fig.~\ref{fig:schematic} consist of multiple quantum dots, and each dot can hold up to one exciton. Therefore, each group of $M$ quantum dots represents an effective site that can accommodate up to $M$ photons, which can then be transported via the photonic structures. 

Fig. \ref{fig:1_by_M} plots the local steady-state currents $J_{12}$, $J_{23}$, and $J_{13}$ as functions of $1/M$ with $\gamma T_{0}=0.5$ and $U=0$, showing clockwise circulation for $t_3/t_1=0.6$, unidirectional flow for $t_3/t_1=1.0$, and counterclockwise circulation for $t_3/t_1=1.4$. The insets of Fig. \ref{fig:1_by_M} show the occupation numbers on the three sites, $n_{1}, n_{2}, n_{3}$, as functions of $1/M$ with the same set of parameters of the main panels. Importantly, while the local occupation numbers of the photons follow $n_{3}<n_{2}<n_{1}$ in all the insets of Fig. \ref{fig:1_by_M}, the local currents exhibit different patterns depending on $t_{3}/t_{1}$. Therefore, the quantum transport of photons is very different from classical transport in the sense that the quantum current may flow opposite to the direction of the density gradient even in the steady state. Our results thus establish the decoupling of the directions of the local currents and those of the local densities in quantum transport of photons. In other words, it is possible to transport photons from a low-density site to a high-density one in the steady state by using a multi-path geometry. The $1/M=0$ results will be discussed in the next section.

\begin{figure}[t]
	\centering
	\includegraphics[width=0.49\textwidth]{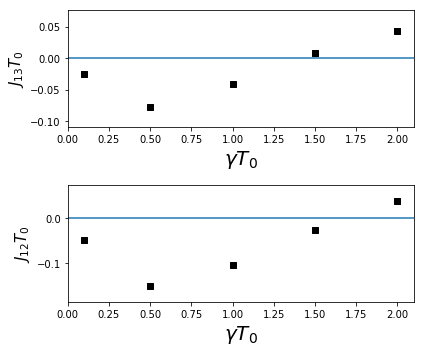}
	\caption{\label{fig:gamma} Tuning the local currents by the system-reservoir coupling $\gamma$: The upper (lower) panel shows $J_{13}$ for $t_{3}/t_{1}=0.6$ and $M=5$  ($J_{12}$ for $t_{3}/t_{1}=1.4$ and $M=5$). Both cases show a change of the sign of the local current as $\gamma$ increases.}
\end{figure}

To further demonstrate the tunability of the local photonic currents, we show the local currents can change signs as $\gamma$ varies. Fig. \ref{fig:gamma} shows $J_{13}$ for fixed $t_3/t_1=0.6$ and $J_{12}$ for fixed $t_3/t_1=1.4$ as functions of $\gamma$, both from the noninteracting case with $M=5$. A change of the sign of the local current implies the reversal of the local-flow direction. The possibility of tuning the local currents using the system-reservoir coupling $\gamma$ introduces additional knobs for controlling the photonic transport. Moreover, the dependence of the local photonic currents on $\gamma$ implies that the photonic circulation results from a combination of the system and reservoirs, so the circulation is not an intrinsic property of the metastructure alone.

We have verified that the results of the cases with asymmetric system-reservoir couplings $(\gamma_{L}\neq \gamma_{R})$ are qualitatively similar to those of the case with the symmetric condition $\gamma_{L} = \gamma_{R}$. All the circulation patterns can be found in the asymmetric cases. We have also checked other values of $N_L$ and $N_R$ of the reservoirs, and the results only differ quantitatively.  Therefore, the photonic circulations are robust against the asymmetry of the system-reservoir couplings and the number of particles in the reservoirs. The robustness of the geometry-induced circulation will make it more feasible to observe the phenomenon in experiments.

\subsection{\label{sec:level3d}Possible applications}
The system shown in Fig.~\ref{fig:schematic} has internal and external parameters. The former includes $t_1$, $t_3$, and $U$, whose values may be determined by the device fabrication or tuned by gate voltage. The latter includes $M$, $\gamma_{L,R}$, and the average number of photons in the reservoirs, which may be tuned by coupling the device to external source or bias. We have shown that both types of parameters can affect the photonic transport. Here we propose two possible applications of the circulating photons in multi-path geometries like the one shown in Fig.~\ref{fig:schematic}. 

The first application is to use the CW and CCW circulations to encode the binary numbers $0$ and $1$ for realizing a memory element. The circulation may be changed by tuning the hopping coefficients, the coupling $\gamma_{L,R}$ with the reservoirs, or the photon blockade $M$ by controlling the underlying quantum dots or the photonic structures and reservoirs. To read out the information, additional photon detectors may couple to the sites and siphon out some photons to measure their momentum, which determines the direction of the photonic current.

The second application follows the possibility of transporting photons from a low-density site to a high-density one using the underlying geometry. The system shown in Fig~\ref{fig:schematic} may be viewed as a proof-of-principle device serving as a controllable local photonic router. If one needs to transport photons from, say, site $2$ to site $1$ regardless of the densities on those sites, one can tune the internal or external parameters to ensure the system stays in the CCW circulation regime. Such a router is made possible by the underlying multi-path geometry, and it is not necessary to have artificial gauge fields or other extra interactions.

\subsection{\label{sec:level3e}Photon circulation without photon blockade} 
Before concluding our work, we show that the photon blockade is not necessary, at least in the noninteracting case, for generating the circulating current in the triangular metastructure shown in Fig.~\ref{fig:schematic} by using the third-quantization formalism \cite{prosen2010quantization} for bosons to find the local currents in the $M\rightarrow \infty$ limit. The formalism is summarized in the Appendix, allowing us to calculate the local flows and densities when the Fock space has the full states without any truncation. We remark that the system-reservoir coupling terms of the Lindblad equation~\eqref{eq:LQME} fit the third-quantization formalism. 

\begin{figure}[t]
	\centering
	\includegraphics[width=0.49\textwidth]{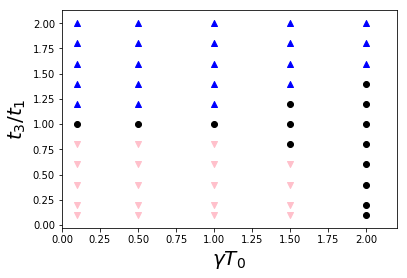}
	\caption{\label{fig:M_inf} Phase diagram showing different steady-state patterns of the photonic current in a noninteracting ($U=0$) system in absence of the photon blockade ($M=\infty$).
	The pink upside-down triangles denote the CW circulation, the black circles denote the UD flow, and the blue triangles denote the CCW circulation.}
\end{figure}

Fig. \ref{fig:M_inf} shows the phase diagram of the steady-state circulations of the noninteracting photons in the metastructure without any limitation of the photon number on each site. All three types of patterns (CW, CCW, UD) are indeed present.
Importantly, the noninteracting photons with $M\rightarrow\infty$ are the genuine case of a noninteracting system because imposing a finite cap $M$ on the number of photons on each site may be considered as introducing effective interactions among the bosons due to the truncated Fock space. Therefore, Fig. \ref{fig:M_inf} establishes two crucial factors of the photon circulation in the triangular metastructure: (1) The photon blockade is not a necessary condition even though experimental setups are likely to introduce the blockade, and (2) the geometry-induced circulation survives in the genuine noninteracting case. 

By comparing Fig. \ref{fig:M_inf} with $M\rightarrow\infty$ to Fig.~\ref{fig:U_0} with $M=1,\cdots, 5$, one can see that the phase diagram without any photon blockade is almost identical to the one with $M=5$.
It is thus expected that the local currents and circulation patterns will remain basically the same as $M$ increases above $5$. Therefore, the circulation regimes saturate with increasing $M$ in absence of the onsite interaction, and it is sufficient to analyze the systems with $M\le 5$ for practical purposes.
The $1/M=0$ results shown earlier in Fig. \ref{fig:1_by_M} were also obtained using the third-quantization approach. As $1/M$ decreases, the local currents and occupation numbers all approach the $1/M=0$ values. The results again suggest that the photonic transport in the triangular metastructure saturates as $M$ increases. One may infer the qualitative behavior of the $M\rightarrow\infty$ limit already in systems with a moderate photon blockade.

The third-quantization method, however, is mainly for noninteracting systems. One may use numerical methods such as the density-matrix renormalization group~\cite{pivzorn2013one,benenti2009charge, lai2016challenges} to include interaction effects in intermediate-sized systems. We have shown that the photon circulation survives in metastrucutres supporting $M\le 5$ photons per site in the presence of the effective photon-photon onsite interactions. There are both theoretical and experimental challenges for realizing and analyzing multi-path metastructures with high photon numbers per site with effective interactions, and they await future research.

\section{\label{sec:level4}Conclusion}
We have demonstrated the possibility of using a multi-path geometry to induce steady-state circulation of local photonic currents without introducing any artificial gauge field. The minimal system for demonstrating the geometry-induced circulation may be realized in a metastructure integrating quantum dots and photonic structures in a triangular geometry. While the photon-exciton interactions may lead to a photon blockade that restricts the number of photons participating in the transport, the photon blockade is not necessary for inducing the circulation, as shown by the third-quantization calculations without any photon blockade. Moreover, the decoupling of the direction of the local current from the density difference between the sites is demonstrated in the metastructure, allowing more tricks for tuning the local transport of photons. Future advances in nanotechnology will allow us to explore more geometric effects in transport phenomena like the one investigated here.

\begin{acknowledgements}
We thank Dr. Chen-Yen Lai for helping with the computer simulations. Part of the simulations were performed on the MERCED cluster funded by National Science Foundation Grant No. ACI-1429783. M.S. acknowledges supported by the Defense Threat Reduction Agency, Basic Research Award No. HDTRA1-15-1-0011.
\end{acknowledgements}

\appendix
\section{Third- quantization formalism}
The local currents and densities with $1/M=0$ in Fig. \ref{fig:1_by_M} and the phase diagrams shown in Fig. \ref{fig:M_inf} were obtained by using the third-quantization formalism for bosons~\cite{prosen2010quantization}. 
The formalism provides an explicit
solution to the dynamics of a bosonic system with a quadratic Hamiltonian described by the Lindblad equation. There is no truncation in the Fock space, so $M\rightarrow\infty$. However, the Lindblad operator needs to be linear in the creation and annihilation operators in order to construct the exact solution. In the following application of the third quantization formalism, we use the notation of Ref. \cite{prosen2010quantization}.

The Lindblad master equation considered in Ref. \cite{prosen2010quantization} is
\begin{eqnarray}\label{eq:PQME}
\frac{\partial\rho(T)}{\partial T} &=& \mathscr{L}\rho(t) \\
&=&  i[\rho,\mathscr{H}] 
+ \sum_{\mu}^{}(2L_{\mu}\rho L_{\mu}^{\dagger} -\{L_{\mu}^{\dagger}L_{\mu},\rho\}). \nonumber
\end{eqnarray}
Here $L_{\mu}$ denotes the Lindblad operators. The generalized decompositions of $\mathscr{H}$ and $L_{\mu}$ for the system shown in Fig.~\ref{fig:schematic} are
\begin{eqnarray}\label{eq:HandL}
\mathscr{H}=\underline{c^{\dagger}}.\boldsymbol{H}\underline{c}, \nonumber \\
L_{\mu}=l_{\mu}.\underline{c}+k_{\mu}.\underline{c^{\dagger}},
\end{eqnarray} 
where
\begin{equation}
\boldsymbol{H}= \left(\begin{array}{ccc} 
0 & -t_1 & -t_3 \\
-t_1 & 0 & -t_1  \\
-t_3 & -t_1 & 0 
\end{array}
\right)
\end{equation}
encodes the information of the system parameters. $\underline{c}$ and $\underline{c^{\dagger}}$ are column vectors of the creation and annihilation operators. 
$l_{\mu}$ and $k_{\mu}$ are column vectors of the coefficients. $\mu$ is the index over the reservoir terms. $l_{\mu}$ and $k_{\mu}$ for our system are given by
\begin{eqnarray}\label{eq:landkmus}
l_{1}= \left(\begin{array}{ccc} 0 & 0 & 0 \end{array}
\right)^T ;k_{1}= \left(\begin{array}{ccc} \sqrt{\gamma N_L/2} & 0 & 0 \end{array}
\right)^T; \nonumber \\
l_{2}=  \left(\begin{array}{ccc} \sqrt{\gamma (N_L+1)/2} & 0 & 0 \end{array}
\right)^T;k_{2}= \left(\begin{array}{ccc} 0 & 0 & 0 \end{array}
\right)^T; \nonumber \\
l_{3}= \left(\begin{array}{ccc} 0 & 0 & 0 \end{array}
\right)^T  ;k_{3}= \left(\begin{array}{ccc} 0 & 0 & \sqrt{\gamma N_R/2} \end{array}
\right)^T; \nonumber \\
l_{4}= \left(\begin{array}{ccc} 0 & 0 & \sqrt{\gamma (N_R+1)/2} \end{array}
\right)^T ;k_{4}= \left(\begin{array}{ccc} 0 & 0 & 0 \end{array}
\right)^T. \nonumber \\
\end{eqnarray} 
With those quantities, we can build the matrices $\boldsymbol{M},\boldsymbol{N},$ and $\boldsymbol{L}$  mentioned in Ref. \cite{prosen2010quantization}. For our systems, they have to following expressions:
\begin{equation}
\boldsymbol{M}=\left(\begin{array}{ccc} 
\gamma (N_L+1)/2 & 0 & 0 \\
0 & 0 & 0  \\
0 & 0 &\gamma (N_R+1)/2 
\end{array}
\right).
\end{equation}
\begin{equation}
\boldsymbol{N}=\left(\begin{array}{ccc} 
\gamma (N_L)/2 & 0 & 0 \\
0 & 0 & 0  \\
0 & 0 &\gamma (N_R)/2 
\end{array}
\right).
\end{equation}
\begin{equation}
\boldsymbol{L}=\left(\begin{array}{ccc} 
0 & 0 & 0 \\
0 & 0 & 0  \\
0 & 0 & 0 
\end{array}
\right).
\end{equation}
Here we assume $\gamma_L=\gamma_R=\gamma$.

The Lindbladian $\mathscr{L}$ may be written in terms of $\underline{b}$, the transformation of ${(c_{A,j},c_{B,j},c_{A,j}^{'},c_{B,j}^{'})}^T$, where $c_{A,j},c_{B,j},c_{A,j}^{'},c_{B,j}^{'}$ represent combinations of the creation and annihilation operators $c_j, c_j^{\dagger}$  at site $j$ and satisfy the almost commutation relations of Ref.~\cite{prosen2010quantization}. Explicitly,
\begin{equation}
    \mathscr{L}=\underline{b}.\boldsymbol{S}\underline{b},
\end{equation}
where 
\begin{equation}
\boldsymbol{S}= \left(\begin{array}{cc} 
0 & -X \\
-X^T & Y
\end{array}
\right).
\end{equation}
Here, $X$ and $Y$ are defined as:
\begin{equation}
X= \frac{1}{2}\left(\begin{array}{cc} 
\iota\boldsymbol{\Bar{H}}-\boldsymbol{\Bar{N}}+\boldsymbol{M} & -2\iota\boldsymbol{K}-\boldsymbol{L}+\boldsymbol{L^T} \\
2\iota\boldsymbol{\Bar{K}}-\boldsymbol{\Bar{L}}+\boldsymbol{\Bar{L^T}} & -\iota\boldsymbol{H}-\boldsymbol{N}+\boldsymbol{\Bar{M}}
\end{array}
\right)
\end{equation}
and 
\begin{equation}
Y= \frac{1}{2}\left(\begin{array}{cc} 
-2\iota\boldsymbol{\Bar{K}}-\boldsymbol{\Bar{L}}-\boldsymbol{\Bar{L^T}} & 2\boldsymbol{N} \\
2\boldsymbol{N^T} & 2\iota\boldsymbol{K}-\boldsymbol{L}-\boldsymbol{L^T}
\end{array}
\right).
\end{equation}

According to Ref. \cite{pivzorn2013one}, the continuous Lyapunov equation $X^TZ+ZX=Y$ in the absence of any coherent pumping term can be simplified as
\begin{equation}\label{eq:LyapunovEqn}
    \Tilde{X}\Tilde{Z}+\Tilde{Z}\Tilde{X^{\dagger}}=\Tilde{Y},
\end{equation}
Where
\begin{equation}
X=\left(\begin{array}{cc} 
\Tilde{X} & 0 \\
0 & \Tilde{X}^* 
\end{array}
\right),
Y=\left(\begin{array}{cc} 
0 & \Tilde{Y} \\
\Tilde{Y} & 0 
\end{array}
\right).
\end{equation}
For our system, using the previously defined $\boldsymbol{M},\boldsymbol{N},$ and $\boldsymbol{L}$ matrices in conjunction with Eqs. (18) and (19) of Ref. \cite{prosen2010quantization}, we can write $\Tilde{X}$ and$\Tilde{Y}$ as:
\begin{equation}
\Tilde{X}= \frac{1}{2}\left(\begin{array}{ccc} 
\gamma/2 & -\iota t_1 &-\iota t_3 \\
-\iota t_1 & 0 &-\iota t_1  \\
-\iota t_3 & -\iota t_1 &\gamma/2 
\end{array}
\right).
\end{equation}
and
\begin{equation}
\Tilde{Y}=\left(\begin{array}{ccc} 
\gamma N_L/2 & 0 & 0 \\
0 & 0 & 0  \\
0 & 0 &\gamma N_R/2 
\end{array}
\right).
\end{equation}
Solving Eq.~\eqref{eq:LyapunovEqn} yields the matrix $\Tilde{Z}$, whose elements are equivalent to the correlation functions $ \Tilde{Z_{lj}}=\langle c_j^{\dagger}c_l \rangle$. The  local currents and occupation numbers can then be calculated from the correlation functions. 

\bibliographystyle{apsrev}

\end{document}